%% file: sample-acmcp.tex
\documentclass[sigconf]{acmart}

\AtBeginDocument{%
  }

\copyrightyear{2025}
\acmYear{2025}
\setcopyright{cc}
\setcctype{by}
\acmConference[CIKM '25]{Proceedings of the 34th ACM International Conference on Information and Knowledge Management}{November 10--14, 2025}{Seoul, Republic of Korea}
\acmBooktitle{Proceedings of the 34th ACM International Conference on Information and Knowledge Management (CIKM '25), November 10--14, 2025, Seoul, Republic of Korea}\acmDOI{10.1145/3746252.3761500}
\acmISBN{979-8-4007-2040-6/2025/11}

\usepackage{xcolor}
\usepackage{subfig}
\usepackage[linesnumbered,ruled,vlined]{algorithm2e}
\usepackage{amsmath}
\usepackage{algorithmic}
\usepackage[T1]{fontenc}

\begin{document}

\title{GOProteinGNN: Leveraging Protein Knowledge Graphs for Protein Representation Learning}



\author{Dan Kalifa}
\orcid{0000-0001-6459-6833}
\affiliation{%
  \department{Faculty of Computer Science}
  \institution{Technion Israel Institute of Technology}
  \city{Haifa}
  \country{Israel}
}
\email{kalifadan@cs.technion.ac.il}

\author{Uriel Singer}
\orcid{0000-0001-8451-8533}
\affiliation{%
  \institution{Meta AI}
  \city{Tel Aviv}
  \country{Israel}
}
\email{urielsinger@meta.com}

\author{Kira Radinsky}
\orcid{0009-0007-7918-2204}
\affiliation{%
  \department{Faculty of Computer Science}
  \institution{Technion Israel Institute of Technology}
  \city{Haifa}
  \country{Israel}
}
\email{kirar@cs.technion.ac.il}

\renewcommand{\shortauthors}{Dan Kalifa, Uriel Singer, and Kira Radinsky}

\acmArticleType{Research}
\acmCodeLink{https://github.com/kalifadan/GOProteinGNN}
\acmDataLink{https://github.com/kalifadan/GOProteinGNN}

\input{0-abstract}




\maketitle

\input{1-intro}

\input{2-related-work}

\input{3-methodologies}

\input{4-experiments}

\input{5-ablations}
\input{6-conclusions}









\bibliographystyle{ACM-Reference-Format}
\balance
\bibliography{main}

\end{document}

%% file: 0-abstract.tex
\begin{abstract}
Proteins are central to biological processes and indispensable for living organisms. Accurate representation of proteins is crucial, especially in drug development.
Recent advances have applied machine learning for unsupervised protein representation learning.
However, these approaches often focus solely on the amino acid sequence of proteins and lack factual knowledge about proteins and their interactions, thus limiting their performance.
In this study, we present GOProteinGNN, a novel architecture that enhances protein language models by integrating protein knowledge graph information during the creation of amino acid level representations. Our approach allows for the integration of information at both the individual amino acid level and the entire protein level, enabling a comprehensive and effective learning process through graph-based learning.
By doing so, we can capture complex relationships and dependencies between proteins and their functional annotations, resulting in more robust and contextually enriched protein representations.
Unlike previous methods, GOProteinGNN uniquely learns the entire protein knowledge graph during training, which allows it to capture broader relational nuances and dependencies beyond mere triplets as done in previous work.
We perform a comprehensive evaluation on several downstream tasks, demonstrating that GOProteinGNN consistently outperforms previous methods, showcasing its effectiveness and establishing it as a state-of-the-art solution for protein representation learning.
We discuss the practical integration of GOProteinGNN in a laboratory setting for lipid nanoparticle-based drug delivery, aiming to bypass the blood-brain barrier and discover novel components, with positive results observed in mice.
\end{abstract}

%% file: 1-intro.tex
\section{Introduction}
\label{sec:intro}

Proteins are fundamental to biology and understanding living systems. Accurately representing them has gained considerable attention, with particular importance for drug development~\cite{Liu2024LargeLM}.
In recent years, there has been a surge of interest in employing deep learning techniques to create unsupervised amino acid representations that can be used to represent a protein~\cite{Rives2019BiologicalSA,Rao2021MSAT}. Prevailing approaches have predominantly focused on unraveling amino acid relationships using text-based methodologies, such as ProtBert~\cite{Elnaggar2022ProtTransTU}. 

Recently, several approaches have attempted to directly create a protein representation by integrating external information about proteins, rather than relying solely on the amino acid representations. 
Those methods, such as KeAP~\cite{Zhou2023ProteinRL}, harness vast protein knowledge graphs. These graphs contain diverse information about interactions the protein participated in, the processes in which it is involved, etc. 
The factual knowledge is conveyed through text descriptions.
Figure~\ref{fig:protein_knowledge_graph} illustrates an example of such a protein knowledge graph, showcasing the intricate relations between proteins and Gene Ontology (GO) terms. 
However, current state-of-the-art (SOTA) methods still encounter challenges in effectively incorporating this extensive information.  

Existing paradigms, at times, reduce the intricate protein knowledge graph to mere triplets (Protein, Relation, GO Term), neglecting the complex tapestry of the knowledge graph~\cite{Zhou2023ProteinRL}. These methods have overlooked the importance of capturing a holistic graph representation. For instance, although they may encode annotations, crucial information such as broader relational nuances and contextual dependencies within the graph may be lost unless the entire graph structure is considered. 
The aforementioned models also incorporate knowledge post-encoder, either via loss objectives or pre-training enrichment~\cite{Zhou2023ProteinRL,Elnaggar2022ProtTransTU}. Language models often struggle to accurately embed knowledge, particularly long-tail knowledge~\cite{Kandpal2022LargeLM}. This limitation hinders their ability to effectively incorporate external knowledge into protein representations during the encoder stage. 
Moreover, these models have predominantly concentrated on either amino acid level learning~\cite{Zhou2023ProteinRL,Elnaggar2022ProtTransTU} or protein level learning~\cite{Zhang2022OntoProteinPP,Lam2023OtterKnowledgeBO}, with none addressing both simultaneously, despite evidence that integrating both improves biological relevance~\cite{HardingLarsen2024ProteinRE}.

In this work, we present GOProteinGNN, a novel architecture that integrates amino acid level and protein level representations within a single framework. Our method addresses existing limitations by comprehensively utilizing the entire protein knowledge graph structure during pre-training, ensuring the capture of intricate relationships. The main technical advancement of GOProteinGNN lies in its ability to capture broader relational nuances and dependencies between protein attributes, which existing models often overlook. In a specific context, when an attribute is connected to another attribute, determining how to embed these connections during protein representation learning is a complex challenge. Compared to GOProteinGNN, which addresses this limitation, KeAP~\cite{Zhou2023ProteinRL} and previous studies have failed to account for a holistic view of the protein's interactions within a biological context, potentially leading to a fragmented understanding of its functional significance.

A key innovation of our approach is the integration of knowledge graph information into protein language models (PLMs) through a novel Graph Neural Network Knowledge Injection (GKI) mechanism. This mechanism centers around the use of the [CLS] token, a special token commonly prepended to input sequences in language models like BERT~\cite{Devlin2019BERTPO} to capture a summary representation of the entire sequence~\cite{Devlin2019BERTPO}. In our model, the [CLS] token is introduced at the start of each amino acid sequence and serves as a global aggregator of information throughout the encoding process.
Initially, the PLM interprets amino acid sequences as coherent "sentences", learning contextual representations for individual amino acids in the early transformer encoder layers. The GKI then enriches these representations by mapping them onto nodes in a protein knowledge graph and propagating relational information from neighboring nodes using a graph-learning method. After this graph-based enrichment, the updated node representation, of the [CLS] token, is passed back into subsequent transformer layers.
Critically, the [CLS] token, which attends to all amino acids during the encoding process, captures a global summary of both the sequence and the graph-enhanced context. As the model alternates between transformer encoding and knowledge graph updates, the [CLS] token continuously absorbs and redistributes high-level information, enabling a richer, more holistic protein representation that incorporates both sequential and factual knowledge.

We perform an empirical evaluation across numerous protein tasks, spanning a broad spectrum of biological domains, comparing to superior methods and achieving SOTA performance across various benchmarks. We assess our method's performance through ablation studies. 
Finally, we demonstrate the real-world applicability of GOProteinGNN through its successful integration into the Targeted
Drug Delivery and Personalized Medicine Laboratory, enabling prediction of blood-brain barrier (BBB) permeability for protein–lipid nanoparticle (LNP) complexes. Experimental validation further demonstrates GOProteinGNN’s practical impact in advancing therapeutic strategies for neurodegenerative diseases.

The contributions of this work are threefold:
(1) We present a novel algorithm for protein representation learning that allows the integration of knowledge graph information into PLMs during the encoder stage. This enables a comprehensive and effective learning process through graph-based learning. 
(2) We introduce the GKI mechanism that enriches amino acid sequence representations with protein knowledge by leveraging the [CLS] token as a global information aggregator, enabling seamless integration of relational graph context into the protein language model.
(3) We present an extensive empirical evaluation of our work over numerous real-world bioinformatics tasks, establishing SOTA results, with a real-world impact through its successful application in the prediction of blood-brain barrier permeability for protein-nanoparticle
complexes. Also, we contribute our code to the community~\footnote{\url{https://github.com/kalifadan/GOProteinGNN}}.

%% file: 2-related-work.tex
\begin{figure*}[t]
\centering
    \subfloat[Protein Knowledge Graph.]
        {
            \includegraphics[width=0.35\textwidth]{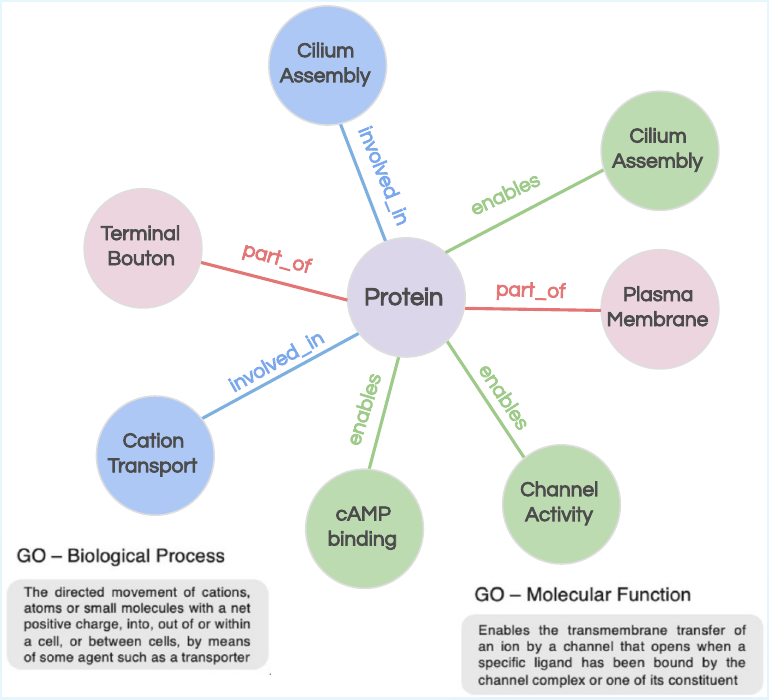}
            \label{fig:protein_knowledge_graph}
        }
    \subfloat[The GKI layer.]
        {
            \includegraphics[width=0.23\textwidth]{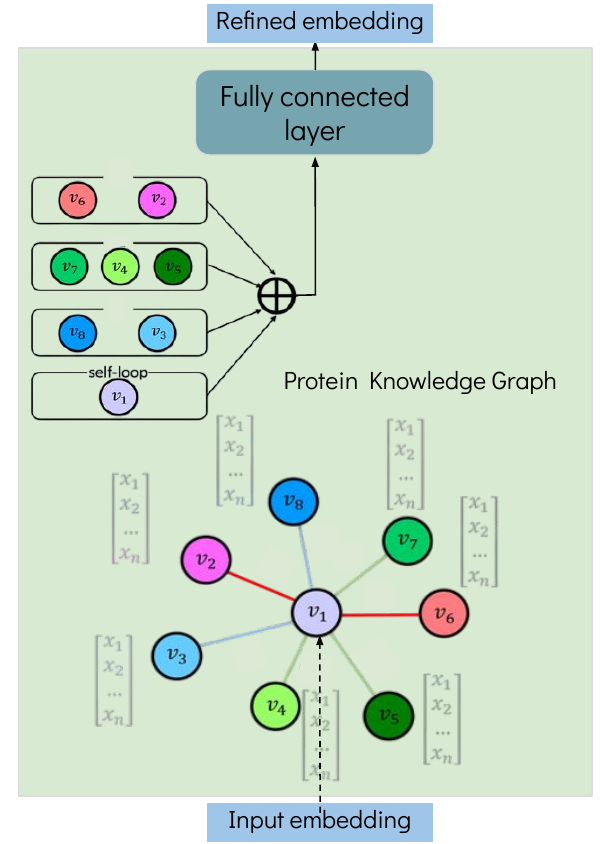}
            \label{fig:gki_layer}
        }
    \subfloat[The GOProteinGNN architecture.]
        {
            \includegraphics[width=0.25\textwidth]{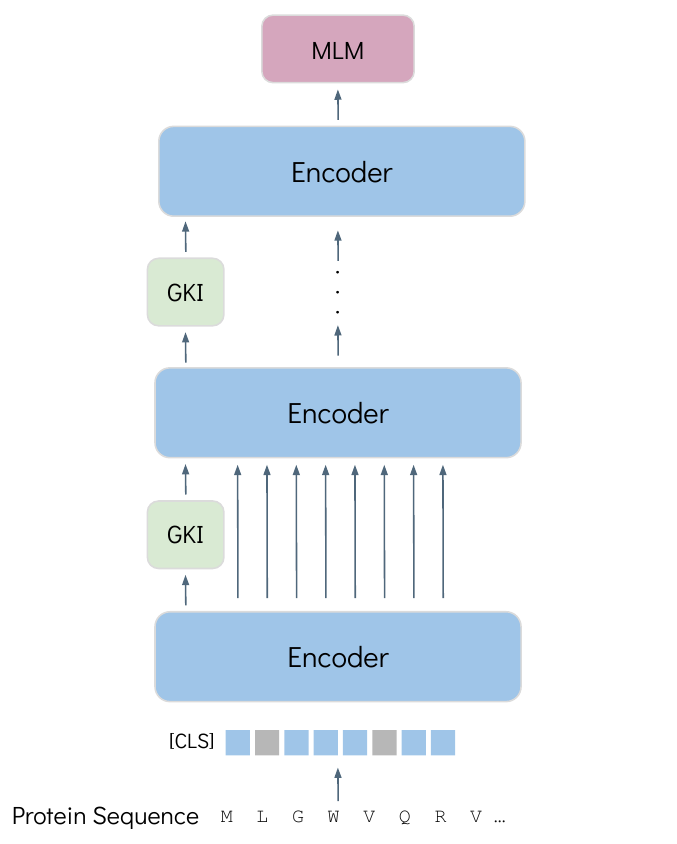}
            \label{fig:GOProteinGNN}
        }
    \caption{\small{The \textbf{GOProteinGNN} pre-training architecture incorporates a protein's knowledge graph with relational and GO term information. First, encoder layers process the amino acid sequence to produce amino acid representations and a protein-level representation (via the [CLS] token). The GKI module then refines the protein representation using graph learning, resulting in a knowledge-enhanced representation that captures key biological context. Finally, subsequent encoder layers further refine the amino acid representations, and the model is trained using masked language modeling (MLM) to predict masked residues.}}
    \label{fig:GOProteinGNN_full}
\end{figure*}

\section{Related Work}
\label{sec:rw}
The protein representation space can be broadly categorized into three approaches: sequence-based methods~\cite{Rives2019BiologicalSA,Lin2022EvolutionaryscalePO,Rao2021MSAT,ma-etal-2024-retrieved}, structure-based methods~\cite{jing2021equivariant,fan2023cdconv,Zhang2022ProteinRL,hermosilla2021ieconv,Wang2022LearningHP}, and structure-informed sequence methods~\cite{Zhang2023ASS,Su2024SaProtPL,hu2024protgo,Wang2022LMGVPAE}. Each category offers distinct strengths and faces unique limitations, depending on the specific task and the availability of structural data.

Sequence-based methods rely on amino acid sequences and draw on advances in natural language processing~\cite{Vaswani2017AttentionIA}. They are highly scalable, trainable on large datasets such as UniRef~\cite{Suzek2007UniRefCA}, and are applicable even when structures are unknown, but they overlook structural information critical for many functions. Structure-based methods capture spatial interactions (e.g., docking) but require limited high-quality structural data and are computationally intensive. Moreover, models that use predicted structures, such as AlphaFold2~\cite{Jumper2021HighlyAP}, may suffer reduced accuracy due to embedding bias between predicted and real structures~\cite{Huang2023Protein3G}.

Within sequence-based methods, a distinct research area focuses on knowledge-enhanced models, which aim to inject structured factual knowledge into language models to improve their reliability and performance~\cite{Sun2019ERNIE2A}.
OntoProtein~\cite{Zhang2022OntoProteinPP} integrates external knowledge graphs into the training process, focusing on GO terms. KeAP~\cite{Zhou2023ProteinRL}, the current SOTA model, explores more granular knowledge graph interactions for improved protein decoding. OntoProtein's contrastive learning and masked modeling are further enhanced by KeAP’s focus on masked language modeling objectives, leading to superior performance across various tasks. 
Beyond proteins, knowledge-enhanced models~\cite{Wang2019KEPLERAU,Lam2023OtterKnowledgeBO,Yao2019KGBERTBF,Liu2019KBERTEL} integrate factual knowledge into language models but lack graph topology modeling and support for diverse node types.

GOProteinGNN is a knowledge-enhanced method within the sequence-based category. While structure-based features can be incorporated into our generic framework, our primary focus remains on knowledge-enhanced modeling. Both KeAP~\cite{Zhou2023ProteinRL} and OntoProtein~\cite{Zhang2022OntoProteinPP} have predominantly focused on either amino acid level learning or protein level learning. They also reduce the intricate protein knowledge graph to mere triplets, neglecting the intricate tapestry of the knowledge graph. GOProteinGNN sets itself apart by uniquely encompassing both levels of representation within a single framework, thus learning the entire protein knowledge graph and allowing it to capture broader relational nuances and dependencies beyond mere triplets. The aforementioned models also incorporate knowledge post-encoder, either via loss objectives or pre-training enrichment~\cite{Zhou2023ProteinRL,Elnaggar2022ProtTransTU}. Language models often struggle to accurately embed knowledge, particularly long-tail knowledge~\cite{Kandpal2022LargeLM}. This limitation hinders their ability to effectively incorporate external knowledge into the representations during the encoder stage.

%% file: 3-methodologies.tex
\section{Methods}
\label{sec:methods}
We present \textbf{GOProteinGNN}, a novel architecture that extends the capabilities of PLMs by integrating structured biological knowledge from protein knowledge graphs. As illustrated in Fig.~\ref{fig:GOProteinGNN_full}, our approach introduces a GKI layer, which enriches amino acid sequence representations with relational information from the knowledge graph. Central to this integration is the use of the [CLS] token—a global representation of the protein sequence—which enables seamless fusion between the PLM and the GKI layer. By alternating between transformer-based sequence modeling and graph-based context propagation, GOProteinGNN produces protein representations that are both contextually rich and biologically informed, capturing essential functional and molecular interactions.

\subsection{GKI Layer}
\label{sec:gki}
We introduce a method called GNN Knowledge Injection (GKI), inspired by GNN~\cite{Scarselli2009TheGN}, a powerful neural network designed for graph-structured data~\cite{Liu2021INDIGOGI,Agiollo2022GNN2GNNGN}. Given a protein knowledge graph, denoted as $G$, and a protein representation, the GKI layer produces an enhanced protein representation that incorporates relevant information from the knowledge graph into the protein representation.

\paragraph{Protein Knowledge Graph}
\label{sec:gki_graph_init}
The protein knowledge graph consists of two different types of nodes: \textit{Proteins} and \textit{GO-Terms}, with \textit{Relations} that act as edges. Proteins are represented as a sequence of amino acids, while the relation and the GO term are terms of factual knowledge described in natural language. The GO terms cover three aspects of biology: Molecular Function, Cellular Component, and Biological Process.
These descriptions are processed through a language encoder~\footnote{We use PubMedBERT \cite{Gu2020DomainSpecificLM}} to generate comprehensive knowledge representations.
To bridge the gap between text and protein, the GKI utilizes affine transformation (linear projection) to project those representations to the same space.
These representations in the protein knowledge graph are denoted by $V_{terms}$.
Let $V_{terms}$ be these representations of GO-terms and $V_{proteins}$ be a set containing proteins (amino acid sequences).
We define the knowledge graph $G = (V, E)$ where:
$V = V_{proteins} \cup V_{terms}$ and $E = \{ (u, v) \mid u \in V_{proteins}, v \in V_{terms}, (u, v) \in Relations \}$.

\paragraph{GNN Method}
Due to the complex diversity of edge types in the protein knowledge graph, we select a method that accommodates this multifaceted edge information. Recognizing that conventional GCNs~\cite{Kipf2016SemiSupervisedCW} lack the capacity to fully exploit such a structure, we turn to the Relational Graph Convolutional Network (RGCN)~\cite{Schlichtkrull2017ModelingRD}, a specialized variant of GNN. This allows us to effectively harness the rich information in the knowledge graph to refine and enhance protein representations, capturing complex relationships and dependencies between proteins and their varying sets of functional annotations, resulting in more contextually enriched representations. While other graph algorithms could be applied, our comparison with more complex methods~\cite{Velickovic2017GraphAN,Zhu2021NeuralBN,galkin2023ultra} showed that RGCN achieves comparable results while being significantly more computationally efficient.
We compare different graph algorithms in Section~\ref{sec:gki_ablation}.

\subsubsection{GKI Procedure}
\label{sec:gki_proc}
Given a protein representation \( z_{protein} \in \mathbb{R}^D \), where $D$ is the feature dimension, corresponding to node $i$ in the knowledge graph $G$, set:
$h_i^{(0)} = z_{protein}$.
The given protein is connected to multiple nodes in the knowledge graph $G$, such as the GO terms nodes ($\subseteq V_{terms}$).
The \( (l+1) \) RGCN layer performs the following update to the node representation \( z_{protein} = h_i^{(l)} \):
\(  h_i^{(l+1)} = \sigma \left( \sum_{r \in R} \sum_{j \in N_r(i)} \frac{1}{c_{i,r}} W_r^{(l)} h_j^{(l)} + W_0^{(l)} h_i^{(l)} \right) \), where \( h_i^{(l+1)} \in \mathbb{R}^D \) is the new node representation, \( \sigma(\cdot) \) is the activation function, \( N_r(i) \) is the set of neighbor nodes of node \( i \) connected by relation type \( r \), \( R \) is the set of all relation types in the graph, \( h_j^{(l)} \in \mathbb{R}^D \) is the node representation of the neighbor node \( j \) at layer \( l \), \( c_{i,r} \) is a normalization term for relation \( r \) specific to node \( i \), and \( W_r^{(l)} \in \mathbb{R}^{D \times D} \) and \( W_0^{(l)} \in \mathbb{R}^{D \times D} \) are learnable weight matrices associated with relation \( r \) and the self-loop, respectively.

The updated protein representation after $L$ RGCN layers is denoted as $z^{'}_{protein} = h_{i}^{(L)}$, capturing essential biological contexts. 

\subsection{GOProteinGNN Algorithm}
\subsubsection{Training Phase}
Following the approach of KeAP \cite{Zhou2023ProteinRL}, we initialize the encoder layers of our model using a large, pre-trained PLM, specifically ProtBert \cite{Elnaggar2022ProtTransTU}. This provides a strong contextual foundation for interpreting amino acid sequences and ensures fair comparison to baseline methods. While we use ProtBert in this work, our method is model-agnostic and could be extended to other advanced PLMs.
During training (See Fig.~\ref{fig:GOProteinGNN_full}), each input protein—represented as an amino acid sequence—is passed through several transformer encoder layers, resulting in intermediate sequence embeddings denoted by $z_p \in \mathbb{R}^{L_p \times D}$, where $L_p$ is the sequence length and $D$ is the embedding dimension.

To represent the entire protein as a single vector, we follow the common practice in transformer models of prepending a special [CLS] token to the sequence \cite{Devlin2019BERTPO}. The [CLS] token is not an amino acid, but a learned embedding that serves as a summary representation of the entire sequence. Throughout the encoding process, this token attends to all amino acids via self-attention, gradually capturing high-level, global information about the protein.
In GOProteinGNN, we repurpose this [CLS] token in a novel way: rather than using it only at the output for classification, we inject structured knowledge from a protein knowledge graph into it multiple times during the encoding process. Specifically, at certain encoder layers, we apply our proposed GKI mechanism to the [CLS] token. At encoder layer $l$, the GKI takes the current $[CLS]'_l$ representation and the protein knowledge graph $G$ as input, producing a knowledge-enriched version of the token $[CLS]'_l$:
$[CLS]'_l = GKI([CLS]_l, G) \in \mathbb{R}^{D}$.
 
This enriched $[CLS]'_l$ token is then passed into the next encoder layer, replacing the original [CLS] representation. Because the [CLS] token interacts with all amino acids through self-attention, it subsequently spreads the injected knowledge across the entire sequence, influencing and refining the representations of each amino acid.

This iterative mechanism enables GOProteinGNN to alternate between sequence-based modeling and knowledge-based enrichment, combining the strengths of both. As a result, the model builds representations that are not only contextually aware but also deeply informed by structured protein knowledge—capturing biological relationships and functional context.
To evenly distribute knowledge integration across the model, we apply the GKI layer $k$ times across $L$ encoder layers, inserting a GKI after every 
$L \div k$ layers. This contrasts to standard PLMs, where the [CLS] token is used only in the final output layer. Instead, GOProteinGNN leverages the [CLS] token throughout the model, ensuring a consistent and dynamic integration of knowledge at multiple stages of the learning.

\begin{table*}[htbp]
\centering
\caption{Results on amino acid contact prediction task. 
Short-range, medium-range, and long-range contacts are contacts between positions that are separated by 6 to 11, 12 to 23, and 24 or more positions, respectively. 
P@L, P@L/2, and P@L/5 denote the precision scores calculated upon top L (i.e., L most likely contacts), top L/2, and top L/5 predictions, respectively. Statistically significant results with \( p < 0.05 \) using a paired t-test are marked with an asterisk (*). The best result is highlighted in bold.}
\renewcommand{\arraystretch}{0.75}
\begin{tabular}{l|ccc|ccc|ccc}
\toprule
Method & \multicolumn{3}{c|}{Short Range} & \multicolumn{3}{c|}{Medium Range} & \multicolumn{3}{c}{Long Range} \\
& P@L & P@L/2 & P@L/5 & P@L & P@L/2 & P@L/5 & P@L & P@L/2 & P@L/5 \\
\midrule
\midrule
LSTM & 0.26 & 0.36 & 0.49 & 0.20 & 0.26 & 0.34 & 0.20 & 0.23 & 0.27 \\
ResNet & 0.25 & 0.34 & 0.46 & 0.28 & 0.25 & 0.35 & 0.10 & 0.13 & 0.17 \\
Transformer & 0.28 & 0.35 & 0.46 & 0.19 & 0.25 & 0.33 & 0.17 & 0.20 & 0.24 \\
ProtBert \cite{Elnaggar2022ProtTransTU} & 0.30 & 0.40 & 0.52 & 0.27 & 0.35 & 0.47 & 0.20 & 0.26 & 0.34 \\
ESM-1b \cite{Rives2019BiologicalSA} & 0.38 & 0.48 & 0.62 & 0.33 & 0.43 & 0.56 & 0.26 & 0.34 & 0.45 \\
ESM-2~\cite{Lin2022EvolutionaryscalePO} & 0.40 & 0.50 & 0.62 & 0.35 & 0.44 & 0.56 & 0.27 & 0.35 & 0.45 \\
KeAP \cite{Zhou2023ProteinRL} & 0.41 & 0.51 & 0.63 & 0.36 & 0.45 & 0.54 & 0.28 & 0.35 & 0.43 \\
\textbf{GOProteinGNN} & $\textbf{0.44}^*$ & $\textbf{0.54}^*$ & $\textbf{0.66}^*$ & $\textbf{0.39}^*$ & $\textbf{0.50}^*$ & $\textbf{0.60}^*$ & $\textbf{0.30}^*$ & $\textbf{0.39}^*$ & $\textbf{0.48}^*$ \\
\bottomrule
\end{tabular}
\label{tab:amino_acid_contact}
\end{table*}

\subsubsection{Masked Language Modeling Training Objective}
For each protein, 20\% of its amino acids are randomly masked. Each masked amino acid $a_{(j)}$ has an 80\% chance of being masked for prediction, 10\% chance of replacement with a random amino acid, and 10\% chance of remaining unchanged. Suppose the number of masked
amino acids is $M$, the training objective $\mathcal{L}_{MLM}$ is to minimize: $\mathcal{L}_{MLM} = -\sum_{j=0}^{M - 1} \log P(a_{(j)} | z_{p}, \Theta)$,
where $z_{p}$ is the protein representation and $\Theta$ are the parameters of the GOProteinGNN model.

\subsubsection{Inference Phase}
During inference, proteins often lack factual knowledge, particularly for unseen cases, and relying on it would risk label leakage. To avoid this dependency, GOProteinGNN omits the GKI components and instead uses a weight-less injection mechanism, with task-specific heads enabling predictions. This design allows the model to generalize across diverse protein tasks without requiring external annotations.

%% file: 4-experiments.tex
\section{Empirical Evaluation}
\label{sec:experiments}

\subsection{Pre-training Dataset}
\label{sec:data_set}
We employed the ProteinKG25 dataset \cite{Zhang2022OntoProteinPP} for pre-training, consistent with all knowledge-enhanced baseline methods, such as OntoProtein~\cite{Zhang2022OntoProteinPP} and KeAP~\cite{Zhou2023ProteinRL}.
ProteinKG25 provides a knowledge graph with aligned descriptions and protein sequences, respectively, to GO terms \cite{Ashburner2000GeneOT} and protein entities.
GO terms can be a molecular function, a cellular component, or a biological process.
The GO terms and their relations to proteins are described using natural language. 
The dataset contains around 5 million triplets, including nearly 600k proteins, 50k attribute terms, and 31 relation terms, with an average of 9 relations per protein.
Consistent with KeAP~\cite{Zhou2023ProteinRL}, we address the data leakage issue in OntoProtein~\cite{Zhang2022OntoProteinPP} by removing amino acid sequences present in downstream tasks from the knowledge graph, ensuring they are unseen during inference. This inductive setup allows for a more accurate assessment of the model’s generalization to unseen proteins. KeAP, after mitigating data leakage, is considered SOTA across all tasks. Given the robust performance demonstrated by KeAP, surpassing methods such as OntoProtein, we focus our comparative analysis on KeAP.
While curated knowledge graphs may contain noise, we found minor GO term inconsistencies (<0.02\%) using lexical and synonym matching~\cite{Mougin2015IdentifyingRA}, which we ignored for consistency. Our main contribution lies in advancing representation learning through a novel knowledge injection approach, building on prior SOTA methods that have shown promising results with existing knowledge graphs.

\subsection{Tasks}
\label{sec:tasks}
We present an extensive empirical evaluation of our method across a wide range of protein-level and amino acid-level representation tasks, including both intrinsic tasks (e.g., semantic similarity inference from PROBE~\cite{Unsal2022LearningFP}) and extrinsic tasks (e.g., protein–protein interactions~\cite{lv2021learning}, contact prediction, and remote homology detection from the TAPE benchmark~\cite{tape}). These experiments demonstrate the value of our representations in real-world bioinformatics applications.
All tasks were divided into specific training, validation, and test splits to ensure that each task tested biologically relevant generalization transferable to real-life scenarios.

\subsection{Baselines}
Incorporating SOTA models as baselines and following KeAP~\cite{Zhou2023ProteinRL} for fair comparison, we used ProtBert~\cite{Elnaggar2022ProtTransTU}, ESM-1b, and ESM-2~\cite{Lin2022EvolutionaryscalePO}, which are considered top-performing models in the absence of factual knowledge.
We also evaluated our method against the most recent and advanced knowledge-driven pre-training model, KeAP~\cite{Zhou2023ProteinRL}.
Additionally, we included task-specific baselines~\cite{Hochreiter1997LongSM,He2015DeepRL,Vaswani2017AttentionIA,Brandes2021ProteinBERTAU,dppi,Li2018DeepNN,Chen2019MultifacetedPI,lv2021learning,Rao2021MSAT,Elnaggar2022ProtTransTU}, known to perform well on the specific tasks.

\subsection{Implementation Details}
GOProteinGNN is trained for 50k steps with a learning rate of 1e-4, weight decay of 0.02, and a batch size of 512 over four GPUs (NVIDIA A40, 48GB Memory each). RGCN layers $L$ are set to 1, and the number of GKI components (referred to as $k$) is set to 3. Our experiments demonstrated that the single-layer RGCN effectively captures most of the critical information, similar to prior studies. We found that adding additional layers did not yield significant gains and, in some cases, introduced noise into the information. For downstream tasks, we follow the SOTA KeAP~\cite{Zhou2023ProteinRL} to ensure a fair comparison. Specifically, we use hyperparameters from GNN-PPI~\cite{lv2021learning} for PPI prediction and PROBE~\cite{Unsal2022LearningFP} for binding affinity and similarity inference. For TAPE~\cite{tape}, we performed a grid search on the validation sets to select the optimal hyperparameters. For full implementation details, refer to the provided code.

\subsubsection{GOProteinGNN Computational Complexity}
Efficient computational handling is essential for graph-based models like GOProteinGNN, especially on large-scale protein knowledge graphs. GOProteinGNN operates at approximately $O(|V| * |E|)$, but several optimizations, such as limiting protein hops, reduce this to $O(|V|)$, comparable to KeAP. During application, the complexity of both algorithms is $O(1)$.
Our setup loads the full receptive graph~\cite{Wang2023AMP} per protein for effective modeling. To support scalability, methods like dynamic caching of receptive graphs~\cite{Wang2023AMP} can accommodate larger graphs. These strategies ensure our method remains practical for large datasets.

\section{Empirical Results}

\subsection{Contact Prediction}
\subsubsection{Task Definition}
Contact prediction is a pairwise amino acid task, where each pair $\langle x_i,x_j\rangle$ of input amino
acids from a sequence (protein) $x$ is assigned to a label $y_{ij} \in \{0, 1\}$, where the label denotes whether the amino acids are in contact ($< 8 $\AA)~\cite{tape}.
The data comes from ProteinNet \cite{proteinnet}, and the evaluation is based on precision.

\subsubsection{Task Result}
In Table~\ref{tab:amino_acid_contact}, we present GOProteinGNN's performance on the contact prediction task compared to six baseline methods. GOProteinGNN significantly outperformed all baseline methods across all categories and metrics, as defined in prior baselines~\cite{Zhou2023ProteinRL,tape}. In particular, GOProteinGNN showed superior performance compared to ProtBert~\cite{Elnaggar2022ProtTransTU} and ESM-2~\cite{Lin2022EvolutionaryscalePO}, which lack knowledge graph information.
This highlights the importance of incorporating knowledge during representation learning, especially for pairwise amino acid tasks, where success depends on understanding the proteins' functions. 
Also, KeAP~\cite{Zhou2023ProteinRL} incorporates a knowledge post-encoder, limiting its capacity to seamlessly integrate external knowledge into protein representations, as reflected in the performance gains of GOProteinGNN over KeAP. This highlights the significant impact of incorporating knowledge graph information in GOProteinGNN via the GKI layers, leading to superior performance in the contact task.

\subsection{Semantic Similarity Inference}
\subsubsection{Task Definition}
This task evaluates how well representation models capture biomolecular functional similarity, focusing on the biological process (BP) category in the GO dataset. We use the semantic similarity measures from \citeauthor{Unsal2022LearningFP} and compute Spearman's rank correlation between representation and GO-based similarities of protein pairs. Higher correlation values indicate better functional similarity representation.

\subsubsection{Task Result}
Table~\ref{tab:semantic_similarity} presents the results of the semantic similarity inference task. GOProteinGNN outperformed all baselines significantly in capturing semantic similarities, achieving an increase of 23.8\% compared to the second-best model.
Biological processes play a crucial role in many drug development processes, making this category particularly important as it pertains directly to drug development. The success in this task underscores GOProteinGNN's efficacy in acquiring enriched representations that encompass the complex biological processes in which proteins are involved, due to the whole protein knowledge graph learning, surpassing the capabilities of alternative approaches. 
Furthermore, KeAP \cite{Zhou2023ProteinRL} showed inferior performance compared to ESM-1b~\cite{Rives2019BiologicalSA}, which does not use factual knowledge. This result is surprising given that both GOProteinGNN and KeAP utilize the same factual knowledge repository, yet KeAP's method of knowledge injection failed to outperform the model that does not use knowledge at all.

\begin{table}[htbp]
\centering
\caption{Results on semantic similarity inference task. The Spearman's rank is the evaluation metric. Statistically significant results with \( p < 0.05 \) using a paired t-test are marked with an asterisk (*). The best result is highlighted in bold.}
\renewcommand{\arraystretch}{0.75}
\begin{tabular}{l|r}
\toprule
Method & Biological Process \\
\midrule
\midrule
MSA Transformer & 0.31 \\
ProtT5-XL & 0.21 \\
ProtBert & 0.35 \\
ESM-1b & 0.42 \\
ESM-2 & 0.41 \\
KeAP & 0.41 \\
\textbf{GOProteinGNN} & $\textbf{0.52}^*$ \\
\bottomrule
\end{tabular}
\label{tab:semantic_similarity}
\end{table}

\subsection{Protein-Protein Interaction Identification}
\subsubsection{Task Definition}
Protein-protein interactions (PPI) refer to physical contacts between two proteins. The goal is to predict the interaction type of each protein pair, out of 7 possible interactions.
The evaluation is conducted on the datasets: SHS27K \cite{Chen2019MultifacetedPI} and SHS148K \cite{Chen2019MultifacetedPI}. Breadth-first search (BFS) is used to generate test sets from the two datasets. F1 score is used as the evaluation metric.

\subsubsection{Task Result}
GOProteinGNN showed strong performance across the board (See Table~\ref{tab:ppi_results}), achieving the highest F1 scores across all datasets with statistical significance, which can be beneficial for drug discovery or personalized medicine.
In particular, GOProteinGNN obtained SOTA results with an increase of approximately 2.1\% on the SHS27K dataset, and 2.7\% on the SHS148K dataset compared to KeAP \cite{Zhou2023ProteinRL}, with statistical significance. 
In conclusion, the results highlight the strengths of our GOProteinGNN, which enables a better biological understanding of the protein, leading to success in predicting interactions between proteins.

\begin{table}[htbp]
\centering
\caption{Results on protein-protein interaction (PPI) identification task. The table displays the results in two datasets: SHS27K and SHS148K. The breadth-first search (BFS) method is used. The F1 score is the evaluation metric. Statistically significant results with \( p < 0.05 \) using a paired t-test are marked with an asterisk (*). The best result is highlighted in bold.}
\renewcommand{\arraystretch}{0.75}
\begin{tabular}{l|c|c}
\toprule
Method & SHS27K & SHS148K \\
\midrule
\midrule
DNN-PPI & 48.09 & 57.40 \\
DPPI & 41.43 & 52.12 \\
PIPR & 44.48 & 61.83  \\
GNN-PPI & 63.81 & 71.37 \\
ProtBert & 70.94 & 70.32 \\
ESM-1b  & 74.92 & 77.49 \\
ESM-2 & 75.05 & 77.19 \\ 
KeAP & 78.58 & 77.22 \\
\textbf{GOProteinGNN} & $\textbf{80.24}^*$ & $\textbf{79.27}^*$ \\
\bottomrule
\end{tabular}
\label{tab:ppi_results}
\end{table}

\subsection{Remote Homology Detection}
\subsubsection{Task Definition}
Remote Homology Detection is a sequence-level classification task. Each input sequence
(protein) $x$ is assigned a label $y \in \{1, ..., 1195\}$ representing different possible protein folds. The data is from DeepSF \cite{Hou2017DeepSFDC} and the evaluation is based on average accuracy on a fold-level holdout set. 

\subsubsection{Task Result}
Despite the challenging nature of the task, due to the large number of potential protein folds, GOProteinGNN significantly outperformed all methods, as can be seen in Table~\ref{tab:remote_homology}, demonstrating its effectiveness in handling the complexities of this task. Specifically, GOProteinGNN achieved an advantage over KeAP \cite{Zhou2023ProteinRL}, further affirming its ability to capture essential biological contexts and interactions which is crucial for predicting the fold structure. 
%
As remote homology detection is a sequence-level classification task, we can see that both KeAP \cite{Zhou2023ProteinRL} and ProtBert \cite{Elnaggar2022ProtTransTU} achieved equal results, suggesting KeAP fails to leverage the factual knowledge in the protein level. 
This task highlights the novelty of our method which allows for the integration of information at both the individual amino acid level and the entire protein level.

\begin{table}[htbp]
\centering
\caption{Results on remote homology detection task. The evaluation metric is accuracy. Statistically significant results with \( p < 0.05 \) using a paired t-test are marked with an asterisk (*). The best result is highlighted in bold.}
\renewcommand{\arraystretch}{0.75}
\begin{tabular}{l|r}
\toprule
Method & Homology \\
\midrule
\midrule
LSTM & 0.26 \\
ResNet & 0.17 \\
Transformer & 0.21 \\
ProtBert & 0.29 \\
ProteinBert & 0.22 \\
ESM-1b & 0.11 \\
ESM-2 & 0.13 \\
KeAP & 0.29 \\
\textbf{GOProteinGNN} & $\textbf{0.32}^*$ \\
\bottomrule
\end{tabular}
\label{tab:remote_homology}
\end{table}

%% file: 5-ablations.tex
\section{Ablation Tests}
\label{sec:ablations}

\subsection{GKI Ablation}
\label{sec:gki_ablation}
The ablation test for the GKI component is presented in Table~\ref{tab:amino_acid_contact_ablation}. The full model results highlight the effectiveness of GOProteinGNN's model components, where the model demonstrated superior performance across all metrics, underscoring its pivotal role.
Ignoring relations through the GKI component implies using GCN \cite{Kipf2016SemiSupervisedCW} instead of RGCN \cite{Schlichtkrull2017ModelingRD} and ignoring edge type during the knowledge graph learning. Without relation learning, we observe a slight decrease in results, emphasizing its importance.
Moreover, excluding the GKI component (i.e., without knowledge integration) substantially degrades performance, reinforcing the significance of the GNN mechanism for effective knowledge integration. Also, employing GAT~\cite{Velickovic2017GraphAN} instead of RGCN achieved comparable results while being significantly more computationally inefficient.
These experiments highlight the contribution of the complete GKI component to GOProteinGNN's ability to capture essential biological contexts and interactions, leading to performance improvements.

\begin{table}[htbp]
\centering
\caption{Ablation tests for the GKI components and the number of its instances ($k$) on the contact prediction task, and the PPI task (SHS27K). The best result is highlighted in bold.}
\resizebox{\columnwidth}{!}{%
\renewcommand{\arraystretch}{0.75}
\begin{tabular}{l|ccc|c}
\toprule
Method & \multicolumn{3}{c|}{Short Range Contact} & \multicolumn{1}{c}{PPI} \\
& P@L & P@L/2 & P@L/5 & F1 \\
\midrule
\midrule
GOProteinGNN & $\textbf{0.44}^*$ & $\textbf{0.54}^*$ & $\textbf{0.66}^*$ & $\textbf{80.24}^*$ \\
$+$ GKI with GAT~\cite{Velickovic2017GraphAN} & 0.43 & 0.53 & 0.65 & 79.96 \\
$-$ GKI Ignoring Relations & 0.41 & 0.51 & 0.64 & 79.14 \\
$-$ Without GKI & 0.30 & 0.40 & 0.52 & 70.94 \\
\midrule
GOProteinGNN ($k = 3$) & $\textbf{0.44}^*$ & $\textbf{0.54}^*$ & $\textbf{0.66}^*$ & $\textbf{80.24}^*$ \\
GOProteinGNN ($k = 2$) & 0.43 & 0.53 & 0.64 & 79.28 \\
\bottomrule
\end{tabular}}
\label{tab:amino_acid_contact_ablation}
\end{table}

\subsection{Number of GKI Components Ablation}
We wish to experiment with the number of optimal GKI components. Each GKI component adds an additional knowledge graph injection layer to the model.
In the case of GOProteinGNN, the number of GKI components (referred to as $k$) is set to 3. In this ablation test, we reduce the number of GKI components to observe their impact.
As shown in Table~\ref{tab:amino_acid_contact_ablation}, reducing knowledge injections in GOProteinGNN results in a slight performance decrease. This decrease highlights the significance and meaningfulness of the GKI component.

%% file: 6-conclusions.tex
\section{Real-World Deployment}
\label{sec:deploy}
GOProteinGNN’s most notable achievement is its deployment at the Technion’s Targeted Drug Delivery Lab, where it played a key role in brain-targeted drug delivery experiments. A major challenge is enabling monoclonal antibodies (mAbs) to cross the BBB for treating Parkinson’s disease, where traditional methods have shown limited success.

\paragraph{Problem Definition.}
The laboratory focuses on developing LNP formulations as delivery vehicles, with an emphasis on improving targeting via protein modifications. Protein selection followed a standard computational screening based on GO annotations, prioritizing those involved in extracellular vesicle transport. This strategy was informed by recent advances in vesicle-based drug delivery, knowledge of BBB transport mechanisms, and prior literature~\cite{Pandit2019TheBB,Sicherman2025ReactEmbedAC}. The final dataset includes over 500 candidate proteins with potential for BBB penetration.

\begin{figure}[h!]
    \centering
    \includegraphics[width=0.34\textwidth]{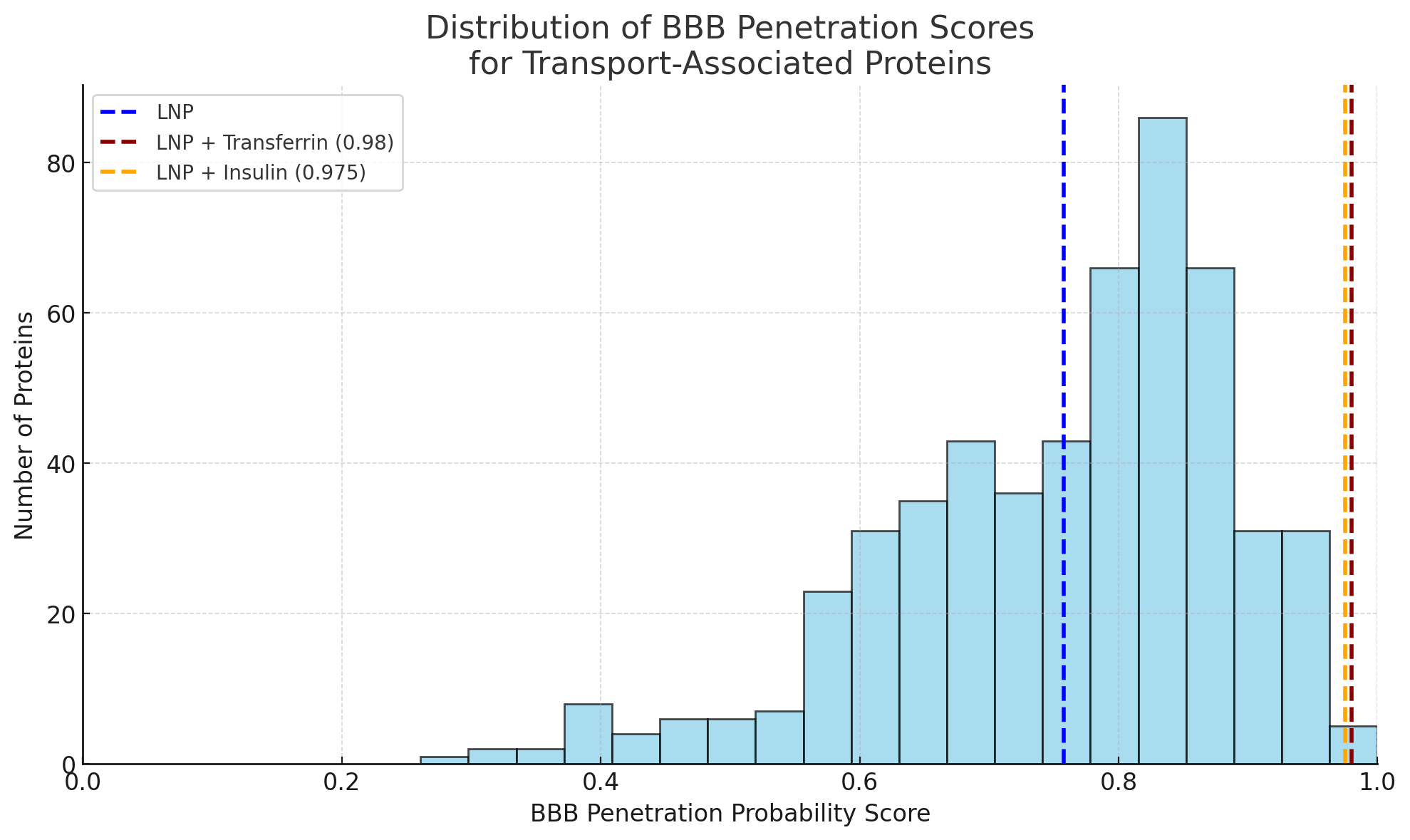}
    \caption{Distribution of BBB Penetration Scores.}
    \label{fig:bbb}
\end{figure}

\paragraph{Methodological Framework.}
We leverage GOProteinGNN to identify optimal proteins for decorating LNPs to enhance brain targeting. To model interactions between small molecules and proteins in a unified representation space, we incorporated a molecular encoder~\cite{Sicherman2025ReactEmbedAC}, alongside GOProteinGNN serving as the protein representation model. We train the molecular encoder on molecular BBBP data (approximately 2,000 molecules) and generating protein representations for all protein candidates. Then, we evaluated the probability of protein-LNP combinations crossing the BBB (See Figure~\ref{fig:bbb}). 
Among the protein candidates, Transferrin and Insulin emerged as the top candidates, with the highest predicted BBB penetration probability above 97\%, an increase of over 22\% compared to the base LNP alone. By encoding external biological knowledge from GO terms, GOProteinGNN enhanced protein representations with BBB-relevant functional signals (e.g., vesicle transport), enabling the identification of novel candidate proteins.

\paragraph{Experimental Assessment.}
Laboratory validation confirmed significant efficacy. Transferrin-functionalized brain-targeted liposomes achieved a sevenfold increase in mAb concentration within in vivo brain cells compared to conventional delivery methods. Additionally, a separate research group independently validated the effectiveness of insulin as a targeting protein, supporting its role in enhancing BBB penetration~\cite{GLADDING2024107938}. The success of GOProteinGNN in this study marks a significant advancement in predicting BBB permeability, serving as a foundational step for future applications in optimizing therapeutic delivery across biological barriers.

\section{Conclusions}
\label{sec:conclusions}
In this paper, we introduced GOProteinGNN, a novel architecture that enhances PLMs by integrating knowledge graph information during amino acid-level representation learning. Specifically, GOProteinGNN proposed a GKI layer that leverages the [CLS] token representation of the PLM during the encoder stage in a novel manner.
Our methodology enables the incorporation of information at both the individual amino acid and protein levels, fostering a comprehensive learning process through graph-based techniques.
Furthermore, GOProteinGNN uniquely learns the entire protein knowledge graph during training, allowing it to capture broader relational nuances and dependencies beyond mere triplets as done in previous work.
GOProteinGNN demonstrates strong performance across diverse bioinformatics tasks, including contact prediction and protein–protein interaction identification, with scalability to larger proteins contingent on the chosen base PLM. We posit that the observed performance improvements arise from a deeper comprehension of proteins themselves and the associated factual knowledge. This understanding allows generalization for never-before-seen proteins, as demonstrated on downstream tasks.
By leveraging both the protein knowledge graph and sequential data modeling, GOProteinGNN effectively propagates information regarding proteins with biological connections. In drug development, it can facilitate virtual screening by prioritizing compounds for binding to specific protein targets, underscoring its potential to provide deeper insights into complex biological phenomena and expediting the drug discovery process.
Ultimately, our findings highlight the crucial role of factual knowledge in protein representation learning, reinforcing the need for structured biological information to enhance PLM-based methods.

\subsection*{GenAI Usage Disclosure}
The authors used ChatGPT (GPT-4) to improve the form and clarity of the text in parts of the manuscript. Generative AI tools were not used for data, code, or experimental design.